\documentclass[letter]{aa} 
\usepackage{graphicx}
\usepackage{txfonts}
\usepackage{natbib}     
\usepackage{xcolor} 
\usepackage{threeparttable} 
\usepackage[colorlinks=true,citecolor=blue]{hyperref} 

\bibpunct{(}{)}{;}{a}{}{,} 

\usepackage[normalem]{ulem}

\begin{document}

   \title{A temperature inversion with atomic iron in the ultra-hot dayside atmosphere of WASP-189b}

   \author{F. Yan\inst{1}
                  \and
          E. Pall\'e\inst{2,3}
                  \and
                  A. Reiners\inst{1}
          \and
          K. Molaverdikhani\inst{4}
          \and
                  N. Casasayas-Barris\inst{2,3}
          \and
          L.~Nortmann\inst{1}
                  \and
          G.~Chen\inst{5}
          \and
          P.~Molli\`ere\inst{4}
          \and
          M.~Stangret\inst{2,3}
                }

  \institute{Institut f\"ur Astrophysik, Georg-August-Universit\"at, Friedrich-Hund-Platz 1, D-37077 G\"ottingen, Germany\\
        \email{fei.yan@uni-goettingen.de}
        \and
        Instituto de Astrof{\'i}sica de Canarias (IAC), Calle V{\'i}a Lactea s/n, E-38200 La Laguna, Tenerife, Spain
\and
Departamento de Astrof{\'i}sica, Universidad de La Laguna, 38026  La Laguna, Tenerife, Spain
\and
Max-Planck-Institut f{\"u}r Astronomie, K{\"o}nigstuhl 17, 69117 Heidelberg, Germany
\and
Key Laboratory of Planetary Sciences, Purple Mountain Observatory, Chinese Academy of Sciences, Nanjing 210023, China
}
        \date{Received 29 April, 2020; accepted 03 July, 2020}


  \abstract
 {Temperature inversion layers are predicted to be present in ultra-hot giant planet atmospheres. Although such inversion layers have recently been observed in several ultra-hot Jupiters, the chemical species responsible for creating the inversion remain unidentified. Here, we present observations of the thermal emission spectrum of an ultra-hot Jupiter, WASP-189b, at high spectral resolution using the HARPS-N spectrograph. Using the cross-correlation technique, we detect a strong \ion{Fe}{i} signal. The detected \ion{Fe}{i} spectral lines are found in emission, which is direct evidence of a temperature inversion in the planetary atmosphere. We further performed a retrieval on the observed spectrum using a forward model with an MCMC approach. When assuming a solar metallicity, the best-fit result returns a temperature of $4320_{-100}^{+120}$\,K at the top of the inversion, which is significantly hotter than the planetary equilibrium temperature (2641\,K). The temperature at the bottom of the inversion is determined as $2200_{-800}^{+1000}$\,K. Such a strong temperature inversion is probably created by the absorption of atomic species like \ion{Fe}{i}.
 } 
   \maketitle

%

\section{Introduction}
~\\
Characterising exoplanet atmospheres is an expanding and  quickly evolving subject within the exoplanet field. One of the key properties of an atmosphere is its temperature structure. Ultra-hot giant planets with equilibrium temperatures ($T_\mathrm{eq}$) larger than 2000\,K  have been proposed to possess a temperature inversion (i.e. a stratospheric layer where temperature increases with altitude), because of the strong absorption of titanium oxide (TiO) and vanadium oxide (VO) in their atmospheres \citep{Hubeny2003,Fortney2008}. 

Earlier searches for temperature inversions and TiO/VO were mostly made at low spectral resolutions. Although there are claims or evidence of inversions and TiO detections \citep[e.g.][]{Knutson2008, Desert2008, Todorov2010,Donovan2010,Sedaghati2017}, many of them are the subject of debate \citep[e.g.][]{Charbonneau2008,Zellem2014,Lowe2014,Schwarz2015,Line2016,Espinoza2019}.
\cite{Haynes2015} observed the thermal emission spectrum of an ultra-hot Jupiter (UHJ),  WASP-33b, and found evidence of a temperature inversion. The existence of an inversion in WASP-33b was later confirmed by \cite{Nugroho2017} with a detection of TiO lines in emission line shape using high-resolution spectroscopy.
A temperature inversion has also been detected in another UHJ, WASP-121b, by \cite{Evans2017}. The latter authors observed the low-resolution thermal emission spectrum of WASP-121b and detected an emission feature of $\mathrm{H_2O}$. However, \cite{Merritt2020} searched for TiO and VO in WASP-121b using high-resolution spectroscopy and reported a non-detection of TiO and VO.
In addition, signs of thermal inversions, which are inferred from excess emission due to CO at the 4.5$\mathrm{\mu}m$ photometry band of the \textit{Spitzer} telescope, have been observed in several UHJs including WASP-18b \citep{Sheppard2017, Arcangeli2018} and WASP-103b \citep{Kreidberg2018}.

Recent theoretical simulations predict that atomic metals like Fe and Mg can exist in the atmosphere of UHJs \citep[e.g.][]{Parmentier2018, Kitzmann2018, Helling2019}, and the absorption of these metal lines at optical and UV wavelengths is capable of producing temperature inversions without the presence of TiO/VO \citep{Lothringer2018}. \cite{Lothringer2019} further show that UHJs orbiting early-type stars have stronger temperature inversions. 
Recent transit observations of several UHJs reveal the presence of various atomic species in their atmospheres, including atomic hydrogen \citep[e.g.][]{Yan2018, Casasayas-Barris2018, Jensen2018} and metal lines of Fe, Mg, Ti, and Ca \citep[e.g.][]{Hoeijmakers2018, Hoeijmakers2019, Casasayas-Barris2019, Cauley2019, Yan2019, Turner2020, Stangret2020, Nugroho2020,Ehrenreich2020,Hoeijmakers2020}.
Very recently, \cite{Pino2020} detected \ion{Fe}{i} lines in emission in the dayside spectrum of KELT-9b, indicating the existence of a temperature inversion in the planetary atmosphere.
For WASP-121b, another UHJ for which a temperature inversion has been detected, several kinds of metals have been discovered in its transmission spectrum \citep{Sing2019, Gibson2020, Ben-Yami2020,Hoeijmakers2020-WASP121}.
The detection of metal lines indicates that metal line absorption could be partially or completely responsible for creating temperature inversions. 

Here, we report a detection of neutral iron (\ion{Fe}{i}) in the dayside thermal emission spectrum of \object{WASP-189b} using high-resolution spectroscopy. The detected \ion{Fe}{i} lines are observed in emission line shape, which is direct evidence of a temperature inversion in the planetary atmosphere.
WASP-189b is a UHJ ($T_\mathrm{eq} = 2641 \pm 34$\,K) orbiting a very bright A-type star \citep{Anderson2018}. \cite{Cauley2020} reported a non-detection of atomic species in its transmission spectrum.

The current paper is organised as follows. In Section 2, we show the observations and data reduction. We present the detection of \ion{Fe}{i} using a cross-correlation technique in Section 3 and the retrieval of the temperature profile in Section 4. Conclusions are provided in Section 5.

%

\section{Observations and data reduction}
We observed \object{WASP-189b} with the HARPS-N high-resolution spectrograph (R $\sim$ 115\,000) mounted on the Telescopio Nazionale \textit{Galileo} over two nights. The observation dates were carefully chosen to be close to the secondary eclipses in order to observe the planetary dayside hemisphere. The observation on the  first night  was after the eclipse while the observation on the second night was before the eclipse. Details on airmass, signal-to-noise ratio (S/N), and number of spectra are given in the observation log presented in Table \ref{obs_log}.

%
\begin{table*}
\caption{Observation logs of HARPS-N observation of \object{WASP-189b}.}             
\label{obs_log}      
\centering                          
\begin{threeparttable}
        \begin{tabular}{c c c c c c c }        
        \hline\hline                 
         &      Date & Airmass change & Phase coverage & Exposure time [s] & $N_\mathrm{spectra}$ & S/N range  \tablefootmark{a} \\     
        \hline                       
  Night-1 & 2019-04-13   & 1.79 -- 1.18 -- 1.59 &  0.533 -- 0.624  & 60 & 247 & 65 -- 90\\ 
  Night-2 & 2019-04-29   & 2.08 -- 1.18 -- 2.19 & 0.384 -- 0.497 & 60 & 297 &  45 -- 80\\                   
\hline                                   
        \end{tabular}
\tablefoot{
\tablefoottext{a}{The S/Ns are measured at $\sim$ 5515 $\mathrm{\AA}$ for wavelength points with a size of 0.01 $\mathrm{\AA}$. }
}
\end{threeparttable}      
\end{table*}

The instrument pipeline (Data Reduction Software) produces order-merged one-dimensional spectra with a wavelength coverage of 383\,nm - 690\,nm. We used these reduced spectra to search for atmospheric signatures. The data from the two nights were reduced separately.
We firstly normalised each individual spectra, shifted them to the Earth's rest frame, and then performed a 5-$\mathrm{\sigma}$ clip on each wavelength bin to remove outliers. We resampled the spectra into a wavelength grid with steps corresponding to $\sim$ 0.6 km\,s$^{-1}$.
The continuum level of HARPS-N spectra exhibits a wavelength-dependent variation, which is mainly attributed to a problem of the atmospheric dispersion corrector \citep{Berdinas2016, Casasayas-Barris2019, Nugroho2020}. Other effects, including blaze function stability, atmospheric extinction, and stellar pulsations, can also introduce continuum variations. Therefore, we implemented the following procedures to correct the continuum variation: averaging all the spectra to obtain a master spectrum; dividing each spectrum by the master spectrum; smoothing the result by a Gaussian function with a large standard deviation of 300 points ($\sim$ 180 km\,s$^{-1}$); and dividing the original spectrum by the smoothed result.
We also tested this method on the order-by-order two-dimensional spectra and used a seven-order polynomial fit instead of the Gaussian smoothing. We found that the final detected \ion{Fe}{i} signal (i.e. the peak cross-correlation signal in Section 3.3) is similar but very slightly ($\sim$ 0.2 $\sigma$) higher than the result using order-merged one-dimensional spectra.

We removed the stellar and telluric absorption lines using the \texttt{SYSREM} algorithm (c.f. Appendix A for details). 
The residual spectra were subsequently aligned into the stellar rest frame by correcting the barycentric Earth's radial velocity (BERV) and the stellar systemic radial velocity ($\mathrm{RV_{sys}}$ = -- 20.82 $\pm$ 0.07 km\,s$^{-1}$, which is measured using our HARPS-N spectra). 
The final residual spectra have values close to 1 and contain the planetary spectral features that are buried in the spectral noises.

We estimated the error of each data point using the following method. We firstly assumed a uniform error for all the data points and ran the \texttt{SYSREM} algorithm with five iterations. Then, for each data point, we calculated the standard deviations of the corresponding wavelength bin (i.e. vertical dimension in Fig.\ref{fig-data-processing}) and the exposure frame (i.e. horizontal dimension in Fig.\ref{fig-data-processing}). The uncertainty of each data point is subsequently assigned as the average of these two standard deviation values.

\section{Detection of \ion{Fe}{i} with cross-correlation method}

\subsection{Modelling the emission spectrum of \ion{Fe}{i}}
\begin{figure}
   \centering
   \includegraphics[width=0.4\textwidth]{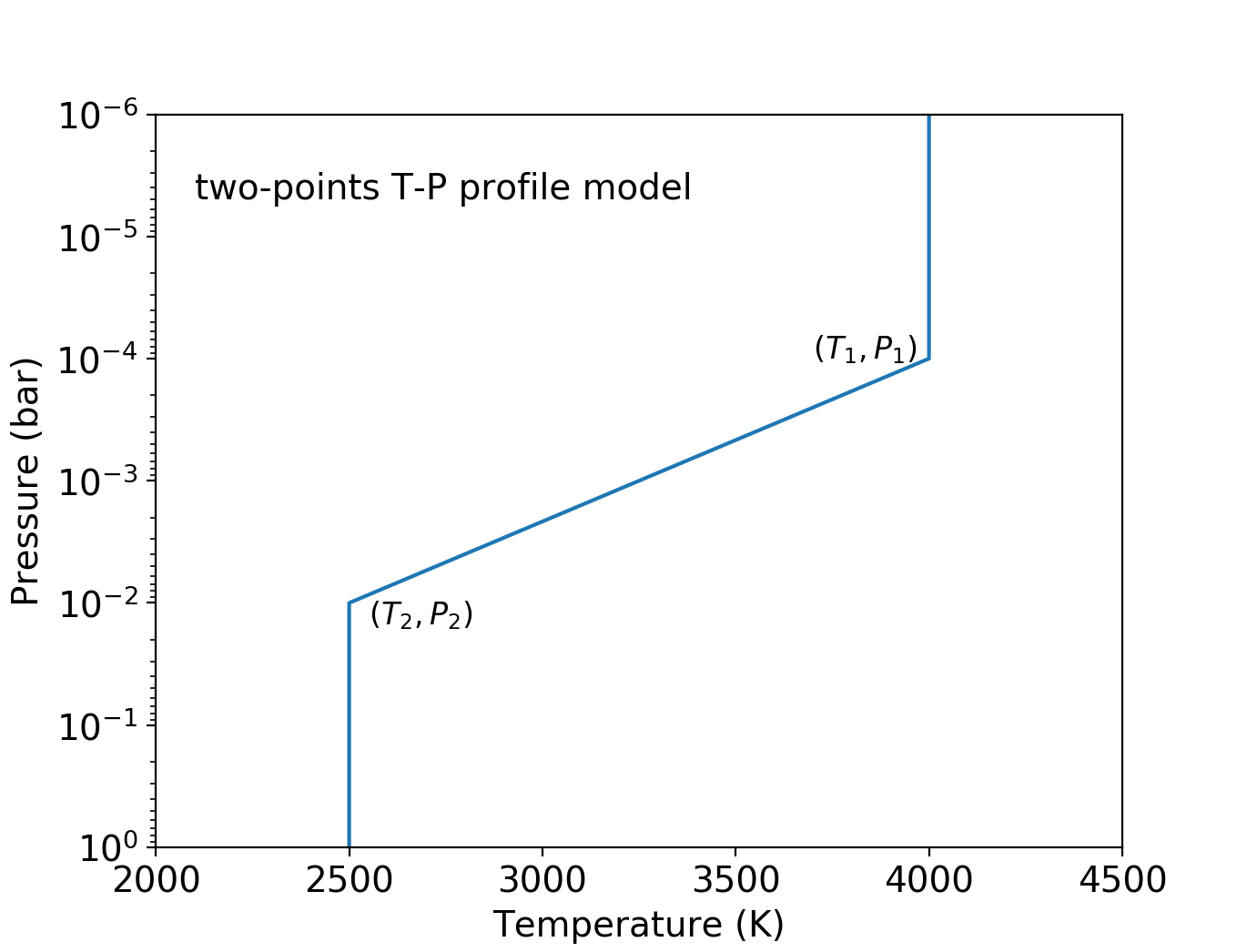}
      \caption{Illustration of the two-point $T$-$P$ profile assumption.}
         \label{fig-TP-example}
   \end{figure}

The cross-correlation method requires a theoretical spectral template to be cross-correlated with the observed spectra \citep{Snellen2010, Brogi2012}. Therefore, we calculated the \ion{Fe}{i} thermal emission spectrum of WASP-189b using the \texttt{petitRADTRANS} code \citep{Molliere2019}. Following the method in \cite{Brogi2014}, we parametrised the atmospheric temperature--pressure ($T$-$P$) profile with a two-point assumption, namely the lower pressure point ($T_\mathrm{1}$, $P_\mathrm{1}$) and the higher pressure point ($T_\mathrm{2}$, $P_\mathrm{2}$).
For pressures lower than $P_\mathrm{1}$ or higher than $P_\mathrm{2}$, the temperatures are assumed to be isothermal; for pressures between $P_\mathrm{1}$ and $P_\mathrm{2}$, the temperature changes linearly with $\mathrm{log_{10}(}P\mathrm{)}$ with a gradient ($T_\mathrm{slope}$) defined as
\begin{equation}
      T_\mathrm{slope} = \frac{ T_\mathrm{1} - T_\mathrm{2} }{ \mathrm{log}P_\mathrm{1} - \mathrm{log}P_\mathrm{2} }.
\end{equation}
For cases with $T_\mathrm{1} > T_\mathrm{2}$, the atmosphere has an inversion layer (i.e. a stratosphere as shown in Fig.\ref{fig-TP-example}).
According to \cite{Lothringer2019}, the $T$-$P$ profiles of UHJs around early-type stars are analogous to the two-point model. 

\begin{figure}
   \centering
   \includegraphics[width=0.48\textwidth]{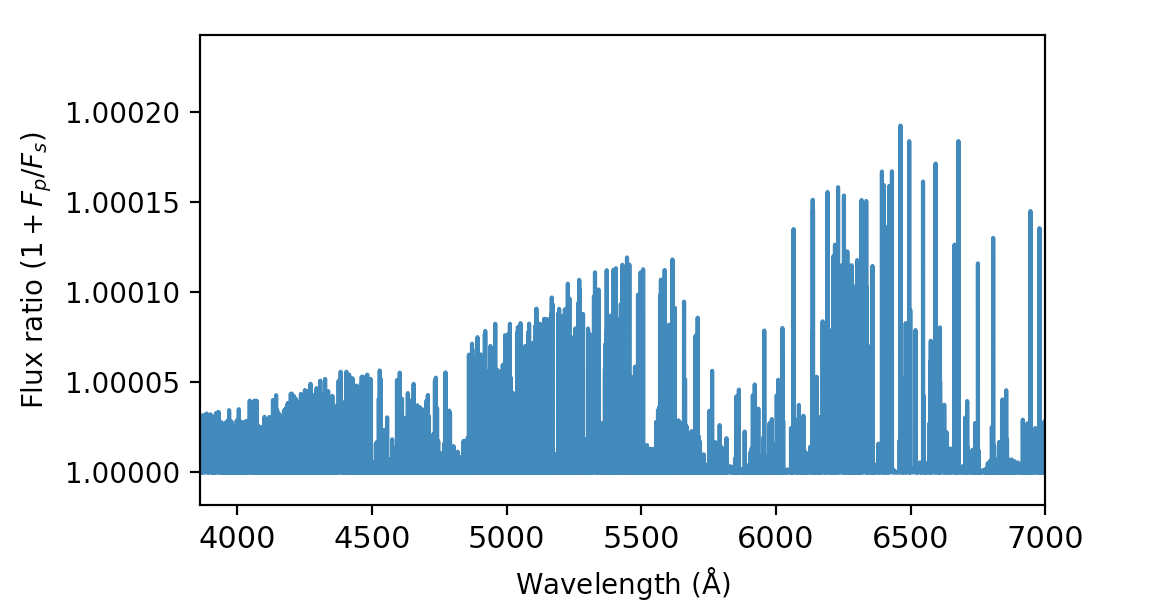}
      \caption{Modelled thermal emission spectrum of \ion{Fe}{i}. We assumed the $T$-$P$ profile as in Fig.\ref{fig-TP-example}.
      The spectrum has been convolved to the resolution of the HARPS-N spectrograph and normalised. This model spectrum is used for cross-correlation.}
         \label{fig-Fe-model-spec}
   \end{figure}

To calculate a model spectrum of WASP-189b, we set the two points as (4000\,K, $10^{-4}$\,bar) and (2500\,K, $10^{-2}$\,bar) (shown in Fig.\ref{fig-TP-example}). We assumed a solar metallicity and set the volume mixing ratio of \ion{Fe}{i} as a constant ($10^{-4.59}$). 
Other parameters including the surface gravity and the stellar and planetary radii ($R_\mathrm{p}$ and $R_\mathrm{s}$) are fixed to the values in \cite{Anderson2018}. 
We neglected $\mathrm{H^-}$ in the model calculation because adding $\mathrm{H^-}$ only  slightly changes the \ion{Fe}{i} line depth  (Fig.\ref{fig-Hminus}).
Rayleigh scattering is also not included in the model.
After obtaining the thermal emission spectrum of the planet ($F_\text{p}$) with \texttt{petitRADTRANS}, we divided it by the blackbody spectrum of the star ($F_\text{s}$) to get the model spectrum 1 + $F_\text{p}/F_\text{s}$. This spectrum was then normalised by dividing it with the continuum spectrum 1 + $C_\text{p}/F_\text{s}$, where $C_\text{p}$ is the blackbody spectrum of $T_\mathrm{2}$.
The spectrum was subsequently convolved with the instrumental profile using the \texttt{broadGaussFast} code from the PyAstronomy library \citep{Czesla2019}. Figure \ref{fig-Fe-model-spec} shows the final model spectrum, in which the  \ion{Fe}{i} lines have emission line shapes because the assumed $T$-$P$ profile has a temperature inversion.

\subsection{Cross-correlation with the \ion{Fe}{i} model}
To search for the \ion{Fe}{i} features in the dayside spectrum, we cross-correlated the observed data (i.e. the residual spectra after \texttt{SYSREM} procedures) with a grid of the model spectrum. The grid was established by shifting the model spectrum from -- 500 km\,s$^{-1}$ to + 500 km\,s$^{-1}$ in 1 km\,s$^{-1}$ steps. 
Here, we did not divide the data by the standard deviation because this procedure would modify the strength of the planetary signal \citep{Brogi2019}. Before the cross-correlation, we filtered the observed residual spectra using a Gaussian high-pass filter with a standard deviation of 15 points ($\sim$ 9 km\,s$^{-1}$) to remove any remaining broad-band features.
For each of the residual spectra, we obtained a cross-correlation function (CCF):
\begin{equation}
      \mathrm{CCF} = \sum r_i m_i ,
\end{equation}
where $r_i$ is the residual spectra and $m_i$ is the shifted model spectrum at wavelength point $i$.
We used the full wavelength coverage of the spectrograph excluding the $\mathrm{O_2}$ band at around 690\,nm and the Na doublet line cores that are affected by interstellar medium absorption.

\subsection{Cross-correlation results}
\begin{figure}
   \centering
   \includegraphics[width=0.5\textwidth]{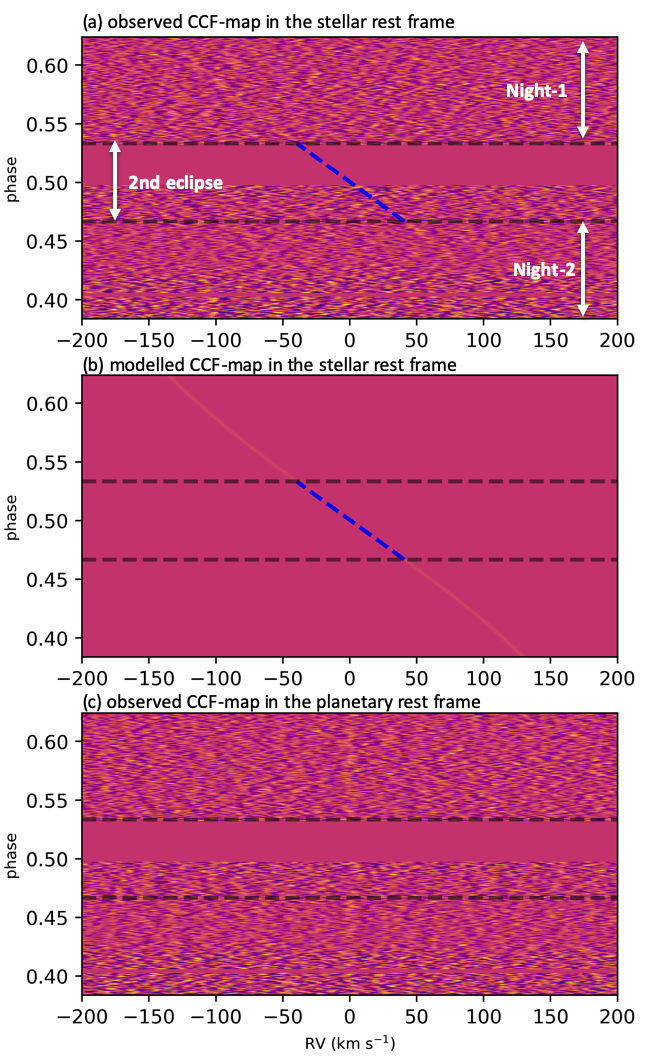}
      \caption{Cross-correlation functions of the observations from the two nights. The upper and middle panels are the observed and modelled CCF map in the stellar rest frame, respectively. The atmospheric \ion{Fe}{i} signal is the bright stripe that appears during out-of-eclipse. The blue dashed lines indicate the expected planetary RV during eclipse.
       The bottom panel is the CCF map with the planetary velocity ($\varv_\mathrm{p}$) corrected using the best-fit  $K_\mathrm{p}$ and $\mathrm{\Delta} \varv$ values, and the atmospheric signal is the vertical stripe located around zero RV.}
         \label{fig-CCF-map}
   \end{figure}

The CCFs of all the observed spectra are presented in Fig. \ref{fig-CCF-map}. The upper panel of Fig. \ref{fig-CCF-map} is the CCF-map in the stellar rest frame (i.e. with BERV and $\mathrm{RV_{sys}}$ corrected). 
The planetary signal, which is the bright stripe with positive radial velocity (RV) before the eclipse and negative RV after the eclipse, is directly observable on the CCF-map. 
We established a model to fit the planetary RV signature directly on the CCF map. We assumed that the CCF has a Gaussian profile and the centre of the Gaussian profile is determined by the planetary velocity:
\begin{equation}
      \varv_\mathrm{p} = K_\mathrm{p} \mathrm{sin(2\pi}\phi) + \mathrm{\Delta} \varv   ,
\end{equation}
where $K_\mathrm{p}$ is the semi-amplitude of the planetary orbital RV, $\phi$ is the orbital phase ($\phi$ = 0 is the mid-transit) and $\mathrm{\Delta} \varv$ is a RV deviation from the planetary orbital velocity. A circular orbit was adopted in the model.
We then applied the Markov Chain Monte Carlo (MCMC) simulations with the \texttt{emcee} tool \citep{Mackey2013} to fit the planetary signal on the CCF map.
For each of the CCFs, we assigned its standard deviation as the noise. The best-fit result is $193.54_{-0.57}^{+0.54}$ km\,s$^{-1}$ for $K_\mathrm{p}$ and $0.66_{-0.26}^{+0.25}$ km\,s$^{-1}$ for $\mathrm{\Delta} \varv$. The obtained $K_\mathrm{p}$ is consistent with the expected $K_\mathrm{p}$ ($197_{-16}^{+15}$ km\,s$^{-1}$) derived using Kepler’s third law with the orbital parameters from \cite{Anderson2018}.
The small $\mathrm{\Delta} \varv$ value could originate from several different sources, including the planetary atmospheric motion due to winds or rotation \citep{Zhang2017,Flowers2019}, the deviation of the absolute stellar RV relative to the measured $\mathrm{RV_{sys}}$ \citep{Yan2019}, the uncertainty in the transit ephemeris (e.g. according to the orbital parameters in \cite{Anderson2018}, the mid-transit time during our observations has an uncertainty of $\sim$ 300\,s, which corresponds to a RV offset of 1.6 km\,s$^{-1}$ at phases close to the secondary eclipse),  and the eccentricity of the orbit.
The lower panel of Fig.\ref{fig-CCF-map} presents the CCF map in the planetary rest frame and the atmospheric signal is the vertical bright stripe located around zero RV, as expected for a planetary signal.

\begin{figure}
   \centering
   \includegraphics[width=0.45\textwidth, height=0.8\textwidth]{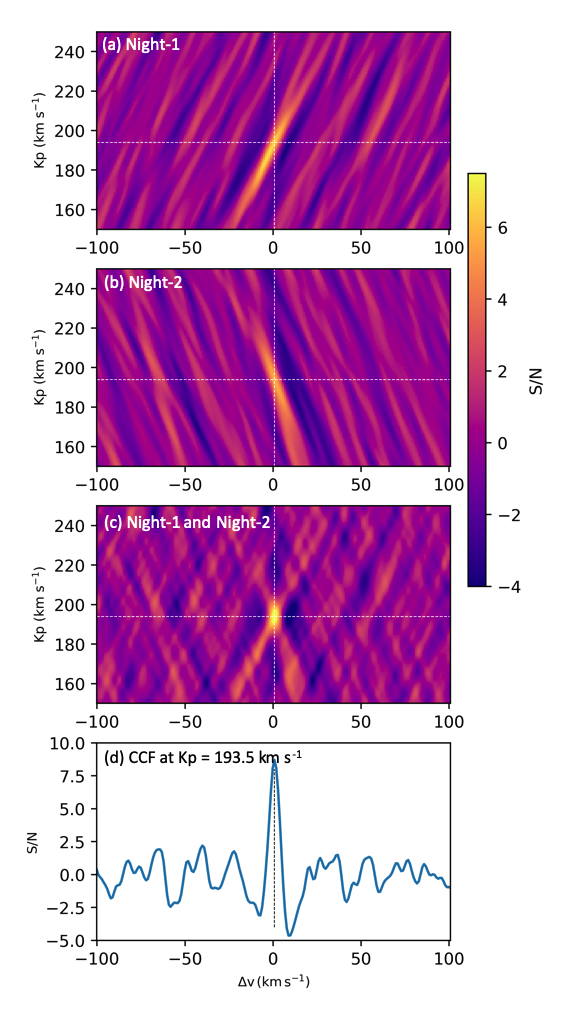}
      \caption{The $K_\mathrm{p}$-$\mathrm{\Delta} \varv$ maps. Panels (a) and (b) are the maps for the Night-1 and Night-2 data, respectively; panel (c) is the map for the data from both nights combined. The bottom panel is the CCF at the best-fit $K_\mathrm{p}$. The dashed lines indicate the best-fit $K_\mathrm{p}$ and $\mathrm{\Delta} \varv$ values.}
         \label{fig-Kp-map}
   \end{figure}

We also generated the classical $K_\mathrm{p}$-map by calculating the combined CCF for different $K_\mathrm{p}$ values (Fig.\ref{fig-Kp-map}). The $K_\mathrm{p}$-map is divided by the standard deviation of the region with $\left| \mathrm{\Delta} \varv \right|$ ranging from 100 to 200 km\,s$^{-1}$. In this way, the value on the $K_\mathrm{p}$-map represents the S/N. 
The Night-1 map has a stronger signal than the Night-2 map because the first half of the Night-2 data have lower S/Ns probably caused by a telescope focusing problem.  The bottom panel presents the combined CCF of the data from the  two nights at $K_\mathrm{p}$ = 193.5 km\,s$^{-1}$, showing a detection of \ion{Fe}{i} at S/N $\sim$ 8.7. The map generated using the data of both nights (Fig.\ref{fig-Kp-map}c) has a very clear peak, demonstrating that $K_\mathrm{p}$ and $\mathrm{\Delta} \varv$ can be well constrained when combining the data observed before and after eclipse.

The strong cross-correlation signal between the observed spectrum and the \ion{Fe}{i} modelled emission spectrum indicates that the planetary \ion{Fe}{i} lines have emission line profiles. Therefore, the detection is clear evidence of the existence of a strong temperature inversion layer in the planetary atmosphere. 
This result agrees with the theoretical simulation in \cite{Lothringer2019}, which suggests that UHJ can have a temperature inversion originating from atomic absorptions.

We also searched for other atomic species including \ion{Fe}{ii}, \ion{Ti}{i,} and \ion{Ti}{ii} using a similar method to that described above (i.e. cross-correlating the observed residual spectrum with the model spectrum assuming a two-point $T$-$P$ profile and constant mixing ratios).
We were not able to detect other atomic species. However, considering that \ion{Fe}{i} has more lines than other atomic species, 
the non-detection does not mean that these species do not exist in the atmosphere. As we are detecting the thermal emission spectrum, other species may exist below or above the temperature inversion where the temperature could be close to isothermal, meaning that these species do not contribute to the emission spectrum.
In addition, we searched for TiO and VO using the cross-correlation method, and did not detect any signal.

\section{Retrieval of the temperature--pressure profile}
~\\
\subsection{Retrieval method}

Although the cross-correlation results demonstrate the existence of a temperature inversion, the detailed $T$-$P$ profile is not constrained. 
Retrieval methods for high-resolution spectroscopy have been developed in recent years \citep[e.g.][]{Brogi2017, Brogi2019, Shulyak2019, Fisher2020, Gibson2020}.
Following these works, we established a framework to retrieve the $T$-$P$ profile of WASP-189b.

Since the $K_\mathrm{p}$ and the $\mathrm{\Delta} v$ are well determined by the cross-correlation method, we fixed these two values and shifted all the residual spectra to the planetary rest frame. We subsequently calculated a master residual spectrum ($R_i$) by averaging all the shifted spectra with their mean $\mathrm{(S/N)^2}$ as weight (Fig.\ref{fig-master-residual}), excluding the ones taken in occultation. 
Following the method in \cite{Gibson2020}, we assumed a standard Gaussian likelihood function and expressed it in logarithm as
\begin{equation}
      \mathrm{ln}(L) = -\frac{1}{2}\sum_{i} \left[ \frac{(R_i - m_i)^2}{(\beta \sigma_i)^2} + \mathrm{ln}(2 \pi (\beta \sigma_i)^2) \right] ,
\end{equation}
where $\sigma_i$ is the uncertainty of the observed residual spectrum $R_i$ at wavelength point $i$ and $\beta$ is a scaling term of the uncertainty. 

The model spectrum $m_i$ is calculated in a similar way as in Section 3.1.1. We assumed a two-points $T$-$P$ profile and set $T_\mathrm{1}$, $P_\mathrm{1}$, $T_\mathrm{2}$, $P_\mathrm{2}$ and $\beta$ as free parameters. We used uniform priors with boundaries shown in Table \ref{tab-mcmc}. In the retrieval work, the atmosphere consists of 81 layers uniformly spaced in log($P$), with a pressure range of $10^{-8}$ bar -- 1 bar.
The volume mixing ratio of \ion{Fe}{i} at each layer is calculated using the chemical module in the \texttt{petitCode} \citep{Molliere2015, Molliere2017}, assuming a solar metallicity and equilibrium chemistry. The model takes into account thermal ionisation and condensation of Fe. Photochemistry is currently not included. According to the simulations of KELT-9b in \cite{Kitzmann2018}, the effect of photochemistry on the \ion{Fe}{i} mixing ratio is relatively weak and is negligible at pressures larger than $10^{-4}$ bar. Therefore, neglecting photochemistry only has a very limited effect on the retrieval.
In addition to the \ion{Fe}{i} lines, we also included \ion{Fe}{ii} in the model to make the retrieval more complete, although the contribution of the \ion{Fe}{ii} lines to the retrieved result is trivial.
The model spectrum is convolved with the instrumental profile and normalised.
We performed the retrieval with the MCMC tool \texttt{emcee}.

\subsection{Retrieval results and discussions}
\begin{figure}
   \centering
   \includegraphics[width=0.45\textwidth]{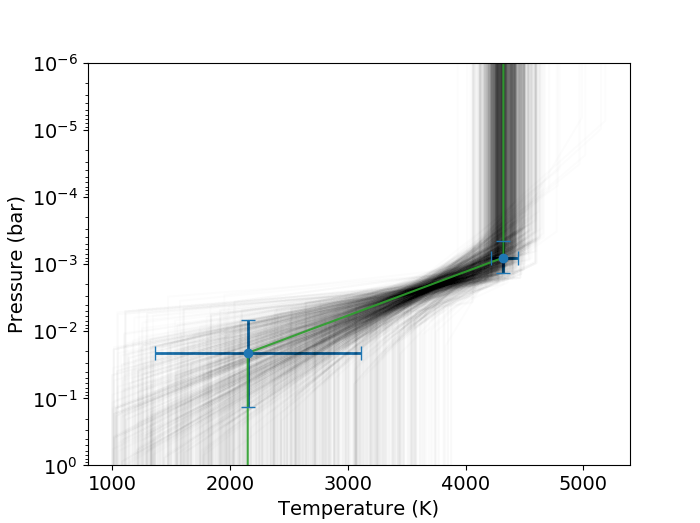}
      \caption{Best-fit $T$-$P$ profile retrieved from the \ion{Fe}{i} emission lines (the blue points and the green line). The grey lines show random examples of $T$-$P$ profile sampled by the MCMC analysis. The result is for the retrieval assuming solar metallicity.}
         \label{fig-best-TP}
   \end{figure}

The best-fit $T$-$P$ profile is presented in Fig.\ref{fig-best-TP} together with a suite of profiles sampled by our analysis. The atmospheric parameters of the best fit are listed in Table \ref{tab-mcmc} and the posterior distributions are                        shown in Fig.\ref{fig-MCMC-TP}. The retrieved $\beta$ parameter has a value close to 1, indicating a proper error estimation in our analysis. 
The retrieved $T$-$P$ profile indicates that the top of the stratosphere is located at $\sim$ $10^{-3.1}$ bar with a temperature of $4320_{-100}^{+120}$ K and the bottom of the stratosphere is located at $\sim$ $10^{-1.7}$ bar with a temperature of $2200_{-800}^{+1000}$ K.  
The contribution functions, which are defined as the fraction of flux a layer contributes to the total flux at given wavelengths, are shown in Fig.\ref{fig-contribution}. According to the contribution functions, the fluxes of the \ion{Fe}{i} emission lines mostly originate from the upper layer of the inversion, and therefore the bottom of the inversion layer is less well confined.
In the wavelength range of HARPS-N, the flux at the line core is determined by $T_\mathrm{1}$ and is much larger than the flux of the adjacent continuum that is determined by the blackbody spectrum at $T_\mathrm{2}$ (c.f. Fig.\ref{fig-continuum}). Therefore, the lower boundary of $T_\mathrm{2}$ cannot be well determined as shown in the posterior distribution plot (Fig.\ref{fig-MCMC-TP}). 
There are other factors that can potentially affect the retrieved values, including uncertainties on the stellar and planetary parameters (e.g. $R_\mathrm{p}/R_\mathrm{s}$) as well as the assumed atmospheric metallicity. 

In the above retrieval work, we assumed the Fe metallicity ([Fe/H]) to be solar, but the retrieved $T$-$P$ profile is related to the assumed [Fe/H]. 
We performed additional retrievals with [Fe/H] = $-1$ (i.e. decreasing the Fe metallicity tenfold) and [Fe/H] = $+1$ (i.e. increasing the Fe metallicity tenfold). The retrieved results are presented in Table \ref{tab-mcmc}.  The $T_\mathrm{1}$ and $T_\mathrm{2}$ values remain similar when changing the Fe metallicity, while the $P_\mathrm{1}$ and $P_\mathrm{2}$ values vary with [Fe/H].
The Fe metallicity is actually degenerate with the pressures. This is because the strength of the \ion{Fe}{i} emission lines are determined by the number density of \ion{Fe}{i} at the upper layer of the inversion. Therefore, when assuming a higher metallicity, the retrieved $P_\mathrm{1}$ value is smaller, resulting in a similar \ion{Fe}{i} number density around ($T_\mathrm{1}$, $P_\mathrm{1}$). 
Detecting other chemical species (e.g. FeH and CO) in the near-infrared will probably allow the degeneracy to be broken and the atmospheric metallicity to be constrained.

The retrieved $T$-$P$ profile is consistent with the prediction by \cite{Lothringer2019} of strong temperature inversions in the atmospheres of UHJs orbiting early-type stars. WASP-189b orbits around an A6-type star and has a very hot dayside atmosphere that is probably dominated by atomic species such as \ion{Fe}{i}. Since the planet receives a large amount of optical and UV irradiation, the absorption of the stellar flux by these atomic species can heat the atmosphere and produce the temperature inversion observed in this work. 
Although our result indicates the inversion could be created by absorptions of atomic species like \ion{Fe}{i}, a comprehensive self-consistent radiation transfer model is required to calculate the actual opacity contributions in the inversion layer.

The strong and dense \ion{Fe}{i} emission lines in the optical wavelengths also increase the overall flux level of the planetary dayside (c.f. Fig.\ref{fig-continuum}). Therefore, when measuring the secondary eclipse in photometry, the obtained eclipse depth will probably be larger than the depth corresponding to the planetary equilibrium temperature.

%
\begin{table*}
\caption{Best-fit parameters from the $T$-$P$ profile retrieval.}             
\label{tab-mcmc}      
\centering                          
\begin{threeparttable}
        \begin{tabular}{c c c c c c}        
        \hline\hline                 
                Parameter & Value ([Fe/H] = 0) & Value ([Fe/H] = $-1$) & Value ([Fe/H] = +1)  & Boundary & Unit \\     
        \hline                       
  \rule{0pt}{2.5ex} $T_\mathrm{1}$ & $4320_{-100}^{+120}$ & $4340_{-100}^{+120}$ & $4330_{-100}^{+130}$ & 1000 to 6000 & K\\ 
  \rule{0pt}{2.5ex} log $P_\mathrm{1}$ & $-3.10_{-0.25}^{+0.23}$ & $-2.36_{-0.20}^{+0.17}$ & $-4.20_{-0.29}^{+0.28}$ & $-7$ to 0 & log bar \\    
  \rule{0pt}{2.5ex} $T_\mathrm{2}$ & $2200_{-800}^{+1000}$ & $2300\pm1000$ & $2100_{-800}^{+900}$ & 1000 to 6000 & K \\ 
  \rule{0pt}{2.5ex} log $P_\mathrm{2}$ & $-1.7_{-0.5}^{+0.8}$ & $-1.3_{-0.3}^{+0.6}$ & $-2.1_{-0.8}^{+1.1}$ & $-7$ to 0 & log bar \\ 
  \rule{0pt}{2.5ex} $\beta$ & $0.871\pm{0.001}$ & $0.871\pm{0.001}$ & $0.871\pm{0.001}$ & 0 to 10 & -\\                 
\hline                                   
        \end{tabular}
\end{threeparttable}      
\end{table*}

\section{Conclusions}
We observed the dayside thermal emission spectrum of WASP-189b with the HARPS-N spectrograph. Utilising the cross-correlation technique, we detected a strong signal of \ion{Fe}{i} lines in emission line shapes, which is direct evidence of a temperature inversion. We fitted the observed \ion{Fe}{i} spectrum with theoretical models assuming a two-point temperature--pressure profile. The retrieved $T$-$P$ profile has a temperature inversion layer located between $\sim$ $10^{-1.7}$ bar and $\sim$ $10^{-3.1}$ bar when assuming a solar metallicity, and the top of the inversion layer has a very hot temperature of $4320_{-100}^{+120}$ K. 
We also searched for other species (including \ion{Fe}{ii}, \ion{Ti}{i}, \ion{Ti}{ii}, and TiO/VO) and were not able to detect them. The non-detection could be attributed to either the lack of sufficient lines in the HARPS-N wavelengths or the non-existence of these species in the temperature inversion layer of the planetary atmosphere.

Several UHJs have temperature inversions detected in their dayside atmospheres, although the formation of the inversion is still under debate. Our detection of a temperature inversion with atomic iron in WASP-189b indicates that temperature inversions in UHJs are probably partially or completed produced by atomic absorptions of the stellar optical and UV irradiation. However, TiO/VO or other metal compounds could still contribute to the formation of temperature inversion. 
Further observations at multiple wavelengths and both low and high spectral resolutions will allow us to decipher comprehensive chemical and temperature structures for these planets.
Expanding the observations to other UHJs will allow us to explore the diversity of temperature structures related to different chemical species and host star types.

\begin{acknowledgements}
We thank the referee for his/her useful comments.
F.Y. acknowledges the support of the DFG priority program SPP 1992 ``Exploring the Diversity of Extrasolar Planets (RE 1664/16-1)''.
This work is based on observations made with the Italian Telescopio Nazionale Galileo (TNG) operated on the island of La Palma by the Fundaci\'on Galileo Galilei of the INAF (Istituto Nazionale di Astrofisica) at the Spanish Observatorio del Roque de los Muchachos of the Instituto de Astrofisica de Canarias. This work is partly financed by the Spanish Ministry of Economics and Competitiveness through project ESP2016-80435-C2-2-R and PGC2018-098153-B-C31.
G.C. acknowledges the support by the Natural Science Foundation of Jiangsu Province (Grant No. BK20190110).
P.M. acknowledges support from the European Research Council under the European Union's Horizon 2020 research and innovation program under grant agreement No. 832428.
\end{acknowledgements}

\bibliographystyle{aa} 
\bibliography{W189-Fe-refer}

\begin{thebibliography}{64}
\expandafter\ifx\csname natexlab\endcsname\relax\def\natexlab#1{#1}\fi

\bibitem[{{Anderson} {et~al.}(2018){Anderson}, {Temple}, {Nielsen}, {Burdanov},
  {Hellier}, {Bouchy}, {Brown}, {Collier Cameron}, {Gillon}, {Jehin}, {Maxted},
  {Pepe}, {Pollacco}, {Pozuelos}, {Queloz}, {S{\'e}gransan}, {Smalley},
  {Triaud}, {Turner}, {Udry}, \& {West}}]{Anderson2018}
{Anderson}, D.~R., {Temple}, L.~Y., {Nielsen}, L.~D., {et~al.} 2018, arXiv
  e-prints, arXiv:1809.04897

\bibitem[{{Arcangeli} {et~al.}(2018){Arcangeli}, {D{\'e}sert}, {Line}, {Bean},
  {Parmentier}, {Stevenson}, {Kreidberg}, {Fortney}, {Mansfield}, \&
  {Showman}}]{Arcangeli2018}
{Arcangeli}, J., {D{\'e}sert}, J.-M., {Line}, M.~R., {et~al.} 2018, \apj, 855,
  L30

\bibitem[{{Ben-Yami} {et~al.}(2020){Ben-Yami}, {Madhusudhan}, {Cabot},
  {Constantinou}, {Piette}, {Gandhi}, \& {Welbanks}}]{Ben-Yami2020}
{Ben-Yami}, M., {Madhusudhan}, N., {Cabot}, S. H.~C., {et~al.} 2020, \apjl,
  897, L5

\bibitem[{{Berdi{\~n}as} {et~al.}(2016){Berdi{\~n}as}, {Amado},
  {Anglada-Escud{\'e}}, {Rodr{\'\i}guez-L{\'o}pez}, \& {Barnes}}]{Berdinas2016}
{Berdi{\~n}as}, Z.~M., {Amado}, P.~J., {Anglada-Escud{\'e}}, G.,
  {Rodr{\'\i}guez-L{\'o}pez}, C., \& {Barnes}, J. 2016, \mnras, 459, 3551

\bibitem[{{Birkby} {et~al.}(2013){Birkby}, {de Kok}, {Brogi}, {de Mooij},
  {Schwarz}, {Albrecht}, \& {Snellen}}]{Birkby2013}
{Birkby}, J.~L., {de Kok}, R.~J., {Brogi}, M., {et~al.} 2013, \mnras, 436, L35

\bibitem[{{Birkby} {et~al.}(2017){Birkby}, {de Kok}, {Brogi}, {Schwarz}, \&
  {Snellen}}]{Birkby2017}
{Birkby}, J.~L., {de Kok}, R.~J., {Brogi}, M., {Schwarz}, H., \& {Snellen},
  I.~A.~G. 2017, \aj, 153, 138

\bibitem[{{Brogi} {et~al.}(2014){Brogi}, {de Kok}, {Birkby}, {Schwarz}, \&
  {Snellen}}]{Brogi2014}
{Brogi}, M., {de Kok}, R.~J., {Birkby}, J.~L., {Schwarz}, H., \& {Snellen},
  I.~A.~G. 2014, \aap, 565, A124

\bibitem[{{Brogi} {et~al.}(2017){Brogi}, {Line}, {Bean}, {D{\'e}sert}, \&
  {Schwarz}}]{Brogi2017}
{Brogi}, M., {Line}, M., {Bean}, J., {D{\'e}sert}, J.~M., \& {Schwarz}, H.
  2017, \apjl, 839, L2

\bibitem[{{Brogi} \& {Line}(2019)}]{Brogi2019}
{Brogi}, M. \& {Line}, M.~R. 2019, \aj, 157, 114

\bibitem[{{Brogi} {et~al.}(2012){Brogi}, {Snellen}, {de Kok}, {Albrecht},
  {Birkby}, \& {de Mooij}}]{Brogi2012}
{Brogi}, M., {Snellen}, I. A.~G., {de Kok}, R.~J., {et~al.} 2012, \nat, 486,
  502

\bibitem[{{Casasayas-Barris} {et~al.}(2018){Casasayas-Barris}, {Pall{\'e}},
  {Yan}, {Chen}, {Albrecht}, {Nortmann}, {Van Eylen}, {Snellen}, {Talens},
  {Gonz{\'a}lez Hern{\'a}ndez}, {Rebolo}, \& {Otten}}]{Casasayas-Barris2018}
{Casasayas-Barris}, N., {Pall{\'e}}, E., {Yan}, F., {et~al.} 2018, \aap, 616,
  A151

\bibitem[{{Casasayas-Barris} {et~al.}(2019){Casasayas-Barris}, {Pall{\'e}},
  {Yan}, {Chen}, {Kohl}, {Stangret}, {Parviainen}, {Helling}, {Watanabe},
  {Czesla}, {Fukui}, {Monta{\~n}{\'e}s-Rodr{\'\i}guez}, {Nagel}, {Narita},
  {Nortmann}, {Nowak}, {Schmitt}, \& {Zapatero Osorio}}]{Casasayas-Barris2019}
{Casasayas-Barris}, N., {Pall{\'e}}, E., {Yan}, F., {et~al.} 2019, \aap, 628,
  A9

\bibitem[{{Cauley} {et~al.}(2019){Cauley}, {Shkolnik}, {Ilyin}, {Strassmeier},
  {Redfield}, \& {Jensen}}]{Cauley2019}
{Cauley}, P.~W., {Shkolnik}, E.~L., {Ilyin}, I., {et~al.} 2019, \aj, 157, 69

\bibitem[{{Cauley} {et~al.}(2020){Cauley}, {Shkolnik}, {Ilyin}, {Strassmeier},
  {Redfield}, \& {Jensen}}]{Cauley2020}
{Cauley}, P.~W., {Shkolnik}, E.~L., {Ilyin}, I., {et~al.} 2020, Research Notes
  of the American Astronomical Society, 4, 53

\bibitem[{{Charbonneau} {et~al.}(2008){Charbonneau}, {Knutson}, {Barman},
  {Allen}, {Mayor}, {Megeath}, {Queloz}, \& {Udry}}]{Charbonneau2008}
{Charbonneau}, D., {Knutson}, H.~A., {Barman}, T., {et~al.} 2008, \apj, 686,
  1341

\bibitem[{{Czesla} {et~al.}(2019){Czesla}, {Schr{\"o}ter}, {Schneider},
  {Huber}, {Pfeifer}, {Andreasen}, \& {Zechmeister}}]{Czesla2019}
{Czesla}, S., {Schr{\"o}ter}, S., {Schneider}, C.~P., {et~al.} 2019, {PyA:
  Python astronomy-related packages}

\bibitem[{{D{\'e}sert} {et~al.}(2008){D{\'e}sert}, {Vidal-Madjar}, {Lecavelier
  Des Etangs}, {Sing}, {Ehrenreich}, {H{\'e}brard}, \& {Ferlet}}]{Desert2008}
{D{\'e}sert}, J.~M., {Vidal-Madjar}, A., {Lecavelier Des Etangs}, A., {et~al.}
  2008, \aap, 492, 585

\bibitem[{{Diamond-Lowe} {et~al.}(2014){Diamond-Lowe}, {Stevenson}, {Bean},
  {Line}, \& {Fortney}}]{Lowe2014}
{Diamond-Lowe}, H., {Stevenson}, K.~B., {Bean}, J.~L., {Line}, M.~R., \&
  {Fortney}, J.~J. 2014, \apj, 796, 66

\bibitem[{Ehrenreich {et~al.}(2020)Ehrenreich, Lovis, Allart, {Zapatero
  Osorio}, Pepe, Cristiani, Rebolo, Santos, Borsa, Demangeon, Dumusque,
  {Gonz{\'{a}}lez Hern{\'{a}}ndez}, Casasayas-Barris, S{\'{e}}gransan, Sousa,
  Abreu, Adibekyan, Affolter, {Allende Prieto}, Alibert, Aliverti, Alves,
  Amate, Avila, Baldini, Bandy, Benz, Bianco, Bolmont, Bouchy, Bourrier, Broeg,
  Cabral, Calderone, Pall{\'{e}}, Cegla, Cirami, Coelho, Conconi, Coretti,
  Cumani, Cupani, Dekker, Delabre, Deiries, D'Odorico, {Di Marcantonio},
  Figueira, Fragoso, Genolet, Genoni, {G{\'{e}}nova Santos}, Hara, Hughes,
  Iwert, Kerber, Knudstrup, Landoni, Lavie, Lizon, Lendl, {Lo Curto}, Maire,
  Manescau, Martins, M{\'{e}}gevand, Mehner, Micela, Modigliani, Molaro,
  Monteiro, Monteiro, Moschetti, M{\"{u}}ller, Nunes, Oggioni, Oliveira,
  Pariani, Pasquini, Poretti, Rasilla, Redaelli, Riva, {Santana Tschudi},
  Santin, Santos, {Segovia Milla}, Seidel, Sosnowska, Sozzetti, Span{\`{o}},
  {Su{\'{a}}rez Mascare{\~{n}}o}, Tabernero, Tenegi, Udry, Zanutta, \&
  Zerbi}]{Ehrenreich2020}
Ehrenreich, D., Lovis, C., Allart, R., {et~al.} 2020, Nature

\bibitem[{{Espinoza} {et~al.}(2019){Espinoza}, {Rackham}, {Jord{\'a}n}, {Apai},
  {L{\'o}pez-Morales}, {Osip}, {Grimm}, {Hoeijmakers}, {Wilson}, {Bixel},
  {McGruder}, {Rodler}, {Weaver}, {Lewis}, {Fortney}, \&
  {Fraine}}]{Espinoza2019}
{Espinoza}, N., {Rackham}, B.~V., {Jord{\'a}n}, A., {et~al.} 2019, \mnras, 482,
  2065

\bibitem[{{Evans} {et~al.}(2017){Evans}, {Sing}, {Kataria}, {Goyal}, {Nikolov},
  {Wakeford}, {Deming}, {Marley}, {Amundsen}, {Ballester}, {Barstow},
  {Ben-Jaffel}, {Bourrier}, {Buchhave}, {Cohen}, {Ehrenreich}, {Garc{\'\i}a
  Mu{\~n}oz}, {Henry}, {Knutson}, {Lavvas}, {Lecavelier Des Etangs}, {Lewis},
  {L{\'o}pez-Morales}, {Mandell}, {Sanz-Forcada}, {Tremblin}, \&
  {Lupu}}]{Evans2017}
{Evans}, T.~M., {Sing}, D.~K., {Kataria}, T., {et~al.} 2017, \nat, 548, 58

\bibitem[{{Fisher} {et~al.}(2020){Fisher}, {Hoeijmakers}, {Kitzmann},
  {M{\'a}rquez-Neila}, {Grimm}, {Sznitman}, \& {Heng}}]{Fisher2020}
{Fisher}, C., {Hoeijmakers}, H.~J., {Kitzmann}, D., {et~al.} 2020, \aj, 159,
  192

\bibitem[{{Flowers} {et~al.}(2019){Flowers}, {Brogi}, {Rauscher}, {Kempton}, \&
  {Chiavassa}}]{Flowers2019}
{Flowers}, E., {Brogi}, M., {Rauscher}, E., {Kempton}, E. M.~R., \&
  {Chiavassa}, A. 2019, \aj, 157, 209

\bibitem[{{Foreman-Mackey} {et~al.}(2013){Foreman-Mackey}, {Hogg}, {Lang}, \&
  {Goodman}}]{Mackey2013}
{Foreman-Mackey}, D., {Hogg}, D.~W., {Lang}, D., \& {Goodman}, J. 2013, \pasp,
  125, 306

\bibitem[{{Fortney} {et~al.}(2008){Fortney}, {Lodders}, {Marley}, \&
  {Freedman}}]{Fortney2008}
{Fortney}, J.~J., {Lodders}, K., {Marley}, M.~S., \& {Freedman}, R.~S. 2008,
  \apj, 678, 1419

\bibitem[{{Gibson} {et~al.}(2020){Gibson}, {Merritt}, {Nugroho}, {Cubillos},
  {de Mooij}, {Mikal-Evans}, {Fossati}, {Lothringer}, {Nikolov}, {Sing},
  {Spake}, {Watson}, \& {Wilson}}]{Gibson2020}
{Gibson}, N.~P., {Merritt}, S., {Nugroho}, S.~K., {et~al.} 2020, \mnras, 493,
  2215

\bibitem[{{Haynes} {et~al.}(2015){Haynes}, {Mandell}, {Madhusudhan}, {Deming},
  \& {Knutson}}]{Haynes2015}
{Haynes}, K., {Mandell}, A.~M., {Madhusudhan}, N., {Deming}, D., \& {Knutson},
  H. 2015, \apj, 806, 146

\bibitem[{{Helling} {et~al.}(2019){Helling}, {Gourbin}, {Woitke}, \&
  {Parmentier}}]{Helling2019}
{Helling}, C., {Gourbin}, P., {Woitke}, P., \& {Parmentier}, V. 2019, \aap,
  626, A133

\bibitem[{{Hoeijmakers} {et~al.}(2020{\natexlab{a}}){Hoeijmakers}, {Cabot},
  {Zhao}, {Buchhave}, {Tronsgaard}, {Kitzmann}, {Grimm}, {Cegla}, {Bourrier},
  {Ehrenreich}, {Heng}, {Lovis}, \& {Fischer}}]{Hoeijmakers2020}
{Hoeijmakers}, H.~J., {Cabot}, S. H.~C., {Zhao}, L., {et~al.}
  2020{\natexlab{a}}, arXiv e-prints, arXiv:2004.08415

\bibitem[{{Hoeijmakers} {et~al.}(2018){Hoeijmakers}, {Ehrenreich}, {Heng},
  {Kitzmann}, {Grimm}, {Allart}, {Deitrick}, {Wyttenbach}, {Oreshenko}, {Pino},
  {Rimmer}, {Molinari}, \& {Di Fabrizio}}]{Hoeijmakers2018}
{Hoeijmakers}, H.~J., {Ehrenreich}, D., {Heng}, K., {et~al.} 2018, \nat, 560,
  453

\bibitem[{{Hoeijmakers} {et~al.}(2019){Hoeijmakers}, {Ehrenreich}, {Kitzmann},
  {Allart}, {Grimm}, {Seidel}, {Wyttenbach}, {Pino}, {Nielsen}, {Fisher},
  {Rimmer}, {Bourrier}, {Cegla}, {Lavie}, {Lovis}, {Patzer}, {Stock}, {Pepe},
  \& {Heng}}]{Hoeijmakers2019}
{Hoeijmakers}, H.~J., {Ehrenreich}, D., {Kitzmann}, D., {et~al.} 2019, \aap,
  627, A165

\bibitem[{{Hoeijmakers} {et~al.}(2020{\natexlab{b}}){Hoeijmakers}, {Seidel},
  {Pino}, {Kitzmann}, {Sindel}, {Ehrenreich}, {Oza}, {Bourrier}, {Allart},
  {Gebek}, {Lovis}, {Yurchenko}, {Astudillo-Defru}, {Bayliss}, {Cegla},
  {Lavie}, {Lendl}, {Melo}, {Murgas}, {Nascimbeni}, {Pepe}, {S{\'e}gransan},
  {Udry}, {Wyttenbach}, \& {Heng}}]{Hoeijmakers2020-WASP121}
{Hoeijmakers}, H.~J., {Seidel}, J.~V., {Pino}, L., {et~al.} 2020{\natexlab{b}},
  arXiv e-prints, arXiv:2006.11308

\bibitem[{{Hubeny} {et~al.}(2003){Hubeny}, {Burrows}, \&
  {Sudarsky}}]{Hubeny2003}
{Hubeny}, I., {Burrows}, A., \& {Sudarsky}, D. 2003, \apj, 594, 1011

\bibitem[{{Jensen} {et~al.}(2018){Jensen}, {Cauley}, {Redfield}, {Cochran}, \&
  {Endl}}]{Jensen2018}
{Jensen}, A.~G., {Cauley}, P.~W., {Redfield}, S., {Cochran}, W.~D., \& {Endl},
  M. 2018, \aj, 156, 154

\bibitem[{{Kitzmann} {et~al.}(2018){Kitzmann}, {Heng}, {Rimmer}, {Hoeijmakers},
  {Tsai}, {Malik}, {Lendl}, {Deitrick}, \& {Demory}}]{Kitzmann2018}
{Kitzmann}, D., {Heng}, K., {Rimmer}, P.~B., {et~al.} 2018, \apj, 863, 183

\bibitem[{{Knutson} {et~al.}(2008){Knutson}, {Charbonneau}, {Allen}, {Burrows},
  \& {Megeath}}]{Knutson2008}
{Knutson}, H.~A., {Charbonneau}, D., {Allen}, L.~E., {Burrows}, A., \&
  {Megeath}, S.~T. 2008, \apj, 673, 526

\bibitem[{{Kreidberg} {et~al.}(2018){Kreidberg}, {Line}, {Parmentier},
  {Stevenson}, {Louden}, {Bonnefoy}, {Faherty}, {Henry}, {Williamson},
  {Stassun}, {Beatty}, {Bean}, {Fortney}, {Showman}, {D{\'e}sert}, \&
  {Arcangeli}}]{Kreidberg2018}
{Kreidberg}, L., {Line}, M.~R., {Parmentier}, V., {et~al.} 2018, \aj, 156, 17

\bibitem[{{Line} {et~al.}(2016){Line}, {Stevenson}, {Bean}, {Desert},
  {Fortney}, {Kreidberg}, {Madhusudhan}, {Showman}, \&
  {Diamond-Lowe}}]{Line2016}
{Line}, M.~R., {Stevenson}, K.~B., {Bean}, J., {et~al.} 2016, \aj, 152, 203

\bibitem[{{Lothringer} \& {Barman}(2019)}]{Lothringer2019}
{Lothringer}, J.~D. \& {Barman}, T. 2019, \apj, 876, 69

\bibitem[{{Lothringer} {et~al.}(2018){Lothringer}, {Barman}, \&
  {Koskinen}}]{Lothringer2018}
{Lothringer}, J.~D., {Barman}, T., \& {Koskinen}, T. 2018, \apj, 866, 27

\bibitem[{{Merritt} {et~al.}(2020){Merritt}, {Gibson}, {Nugroho}, {de Mooij},
  {Hooton}, {Matthews}, {McKemmish}, {Mikal-Evans}, {Nikolov}, {Sing}, {Spake},
  \& {Watson}}]{Merritt2020}
{Merritt}, S.~R., {Gibson}, N.~P., {Nugroho}, S.~K., {et~al.} 2020, \aap, 636,
  A117

\bibitem[{{Molli{\`e}re} {et~al.}(2017){Molli{\`e}re}, {van Boekel}, {Bouwman},
  {Henning}, {Lagage}, \& {Min}}]{Molliere2017}
{Molli{\`e}re}, P., {van Boekel}, R., {Bouwman}, J., {et~al.} 2017, \aap, 600,
  A10

\bibitem[{{Molli{\`e}re} {et~al.}(2015){Molli{\`e}re}, {van Boekel},
  {Dullemond}, {Henning}, \& {Mordasini}}]{Molliere2015}
{Molli{\`e}re}, P., {van Boekel}, R., {Dullemond}, C., {Henning}, T., \&
  {Mordasini}, C. 2015, \apj, 813, 47

\bibitem[{{Molli{\`e}re} {et~al.}(2019){Molli{\`e}re}, {Wardenier}, {van
  Boekel}, {Henning}, {Molaverdikhani}, \& {Snellen}}]{Molliere2019}
{Molli{\`e}re}, P., {Wardenier}, J.~P., {van Boekel}, R., {et~al.} 2019, \aap,
  627, A67

\bibitem[{{Nugroho} {et~al.}(2020){Nugroho}, {Gibson}, {de Mooij}, {Watson},
  {Kawahara}, \& {Merritt}}]{Nugroho2020}
{Nugroho}, S.~K., {Gibson}, N.~P., {de Mooij}, E. J.~W., {et~al.} 2020, \mnras,
  496, 504

\bibitem[{{Nugroho} {et~al.}(2017){Nugroho}, {Kawahara}, {Masuda}, {Hirano},
  {Kotani}, \& {Tajitsu}}]{Nugroho2017}
{Nugroho}, S.~K., {Kawahara}, H., {Masuda}, K., {et~al.} 2017, \aj, 154, 221

\bibitem[{{O'Donovan} {et~al.}(2010){O'Donovan}, {Charbonneau}, {Harrington},
  {Madhusudhan}, {Seager}, {Deming}, \& {Knutson}}]{Donovan2010}
{O'Donovan}, F.~T., {Charbonneau}, D., {Harrington}, J., {et~al.} 2010, \apj,
  710, 1551

\bibitem[{{Parmentier} {et~al.}(2018){Parmentier}, {Line}, {Bean}, {Mansfield},
  {Kreidberg}, {Lupu}, {Visscher}, {D{\'e}sert}, {Fortney}, {Deleuil},
  {Arcangeli}, {Showman}, \& {Marley}}]{Parmentier2018}
{Parmentier}, V., {Line}, M.~R., {Bean}, J.~L., {et~al.} 2018, \aap, 617, A110

\bibitem[{{Pino} {et~al.}(2020){Pino}, {D{\'e}sert}, {Brogi}, {Malavolta},
  {Wyttenbach}, {Line}, {Hoeijmakers}, {Fossati}, {Bonomo}, {Nascimbeni},
  {Panwar}, {Affer}, {Benatti}, {Biazzo}, {Bignamini}, {Borsa}, {Carleo},
  {Claudi}, {Cosentino}, {Covino}, {Damasso}, {Desidera}, {Giacobbe},
  {Harutyunyan}, {Lanza}, {Leto}, {Maggio}, {Maldonado}, {Mancini}, {Micela},
  {Molinari}, {Pagano}, {Piotto}, {Poretti}, {Rainer}, {Scandariato},
  {Sozzetti}, {Allart}, {Borsato}, {Bruno}, {Fabrizio}, {Ehrenreich},
  {Fiorenzano}, {Frustagli}, {Lavie}, {Lovis}, {Magazz{\`u}}, {Nardiello},
  {Pedani}, \& {Smareglia}}]{Pino2020}
{Pino}, L., {D{\'e}sert}, J.-M., {Brogi}, M., {et~al.} 2020, \apjl, 894, L27

\bibitem[{{Schwarz} {et~al.}(2015){Schwarz}, {Brogi}, {de Kok}, {Birkby}, \&
  {Snellen}}]{Schwarz2015}
{Schwarz}, H., {Brogi}, M., {de Kok}, R., {Birkby}, J., \& {Snellen}, I. 2015,
  \aap, 576, A111

\bibitem[{{Sedaghati} {et~al.}(2017){Sedaghati}, {Boffin}, {MacDonald},
  {Gandhi}, {Madhusudhan}, {Gibson}, {Oshagh}, {Claret}, \&
  {Rauer}}]{Sedaghati2017}
{Sedaghati}, E., {Boffin}, H. M.~J., {MacDonald}, R.~J., {et~al.} 2017, \nat,
  549, 238

\bibitem[{{Sheppard} {et~al.}(2017){Sheppard}, {Mandell}, {Tamburo}, {Gand hi},
  {Pinhas}, {Madhusudhan}, \& {Deming}}]{Sheppard2017}
{Sheppard}, K.~B., {Mandell}, A.~M., {Tamburo}, P., {et~al.} 2017, \apjl, 850,
  L32

\bibitem[{{Shulyak} {et~al.}(2019){Shulyak}, {Rengel}, {Reiners}, {Seemann}, \&
  {Yan}}]{Shulyak2019}
{Shulyak}, D., {Rengel}, M., {Reiners}, A., {Seemann}, U., \& {Yan}, F. 2019,
  \aap, 629, A109

\bibitem[{{Sing} {et~al.}(2019){Sing}, {Lavvas}, {Ballester}, {Lecavelier des
  Etangs}, {Marley}, {Nikolov}, {Ben-Jaffel}, {Bourrier}, {Buchhave}, {Deming},
  {Ehrenreich}, {Mikal-Evans}, {Kataria}, {Lewis}, {L{\'o}pez-Morales},
  {Garc{\'\i}a Mu{\~n}oz}, {Henry}, {Sanz-Forcada}, {Spake}, {Wakeford}, \&
  {PanCET Collaboration}}]{Sing2019}
{Sing}, D.~K., {Lavvas}, P., {Ballester}, G.~E., {et~al.} 2019, \aj, 158, 91

\bibitem[{{Snellen} {et~al.}(2014){Snellen}, {Brandl}, {de Kok}, {Brogi},
  {Birkby}, \& {Schwarz}}]{Snellen2014}
{Snellen}, I. A.~G., {Brandl}, B.~R., {de Kok}, R.~J., {et~al.} 2014, \nat,
  509, 63

\bibitem[{{Snellen} {et~al.}(2010){Snellen}, {de Kok}, {de Mooij}, \&
  {Albrecht}}]{Snellen2010}
{Snellen}, I.~A.~G., {de Kok}, R.~J., {de Mooij}, E.~J.~W., \& {Albrecht}, S.
  2010, \nat, 465, 1049

\bibitem[{{Stangret} {et~al.}(2020){Stangret}, {Casasayas-Barris}, {Pall{\'e}},
  {Yan}, {S{\'a}nchez-L{\'o}pez}, \& {L{\'o}pez-Puertas}}]{Stangret2020}
{Stangret}, M., {Casasayas-Barris}, N., {Pall{\'e}}, E., {et~al.} 2020, \aap,
  638, A26

\bibitem[{{Tamuz} {et~al.}(2005){Tamuz}, {Mazeh}, \& {Zucker}}]{Tamuz2005}
{Tamuz}, O., {Mazeh}, T., \& {Zucker}, S. 2005, \mnras, 356, 1466

\bibitem[{{Todorov} {et~al.}(2010){Todorov}, {Deming}, {Harrington},
  {Stevenson}, {Bowman}, {Nymeyer}, {Fortney}, \& {Bakos}}]{Todorov2010}
{Todorov}, K., {Deming}, D., {Harrington}, J., {et~al.} 2010, \apj, 708, 498

\bibitem[{{Turner} {et~al.}(2020){Turner}, {de Mooij}, {Jayawardhana}, {Young},
  {Fossati}, {Koskinen}, {Lothringer}, {Karjalainen}, \&
  {Karjalainen}}]{Turner2020}
{Turner}, J.~D., {de Mooij}, E. J.~W., {Jayawardhana}, R., {et~al.} 2020,
  \apjl, 888, L13

\bibitem[{{Yan} {et~al.}(2019){Yan}, {Casasayas-Barris}, {Molaverdikhani},
  {Alonso-Floriano}, {Reiners}, {Pall{\'e}}, {Henning}, {Molli{\`e}re}, {Chen},
  {Nortmann}, {Snellen}, {Ribas}, {Quirrenbach}, {Caballero}, {Amado},
  {Azzaro}, {Bauer}, {Cort{\'e}s Contreras}, {Czesla}, {Khalafinejad}, {Lara},
  {L{\'o}pez-Puertas}, {Montes}, {Nagel}, {Oshagh}, {S{\'a}nchez-L{\'o}pez},
  {Stangret}, \& {Zechmeister}}]{Yan2019}
{Yan}, F., {Casasayas-Barris}, N., {Molaverdikhani}, K., {et~al.} 2019, \aap,
  632, A69

\bibitem[{{Yan} \& {Henning}(2018)}]{Yan2018}
{Yan}, F. \& {Henning}, T. 2018, Nature Astronomy, 2, 714

\bibitem[{{Zellem} {et~al.}(2014){Zellem}, {Lewis}, {Knutson}, {Griffith},
  {Showman}, {Fortney}, {Cowan}, {Agol}, {Burrows}, {Charbonneau}, {Deming},
  {Laughlin}, \& {Langton}}]{Zellem2014}
{Zellem}, R.~T., {Lewis}, N.~K., {Knutson}, H.~A., {et~al.} 2014, \apj, 790, 53

\bibitem[{{Zhang} {et~al.}(2017){Zhang}, {Kempton}, \& {Rauscher}}]{Zhang2017}
{Zhang}, J., {Kempton}, E. M.~R., \& {Rauscher}, E. 2017, \apj, 851, 84

\end{thebibliography}

\appendix
\section{Removal of stellar and telluric lines with SYSREM}
The \texttt{SYSREM} algorithm \citep{Tamuz2005} has been demonstrated to be a robust technique to remove the stellar and telluric lines for high-resolution spectroscopy \citep{Birkby2013,Snellen2014,Birkby2017}.
The input data for the \texttt{SYSREM} algorithm is the normalised and blaze-variation-corrected spectral matrix. We chose a \texttt{SYSREM} iteration of 6 because the S/Ns of the planetary signal are relatively high for both nights at this iteration. We tested different iterations and found that the results do not change significantly. 

In order to preserve the relative depths of the planetary lines, we used the same method as in \cite{Gibson2020}. Firstly, we applied the \texttt{SYSREM} iterations in the classical way. Subsequently, instead of using the subtracting residual as the final product, we summed the \texttt{SYSREM} models from each iteration and divided the original input data by the final model. In this way, the strengths of the planetary signals located at the stellar and telluric absorption lines are preserved and the residual spectra represent the normalised spectra of 1 + $F_\mathrm{p}/F_\mathrm{s}$. 
We performed the \texttt{SYSREM} with data in flux-space, while the \texttt{SYSREM} method was originally used for data in magnitude-space \citep{Tamuz2005}. We note that dividing the \texttt{SYSREM} models in flux-space as described above is mostly equivalent to subtracting the \texttt{SYSREM} models in magnitude-space.

   \begin{figure}
   \centering
   \includegraphics[width=0.49\textwidth]{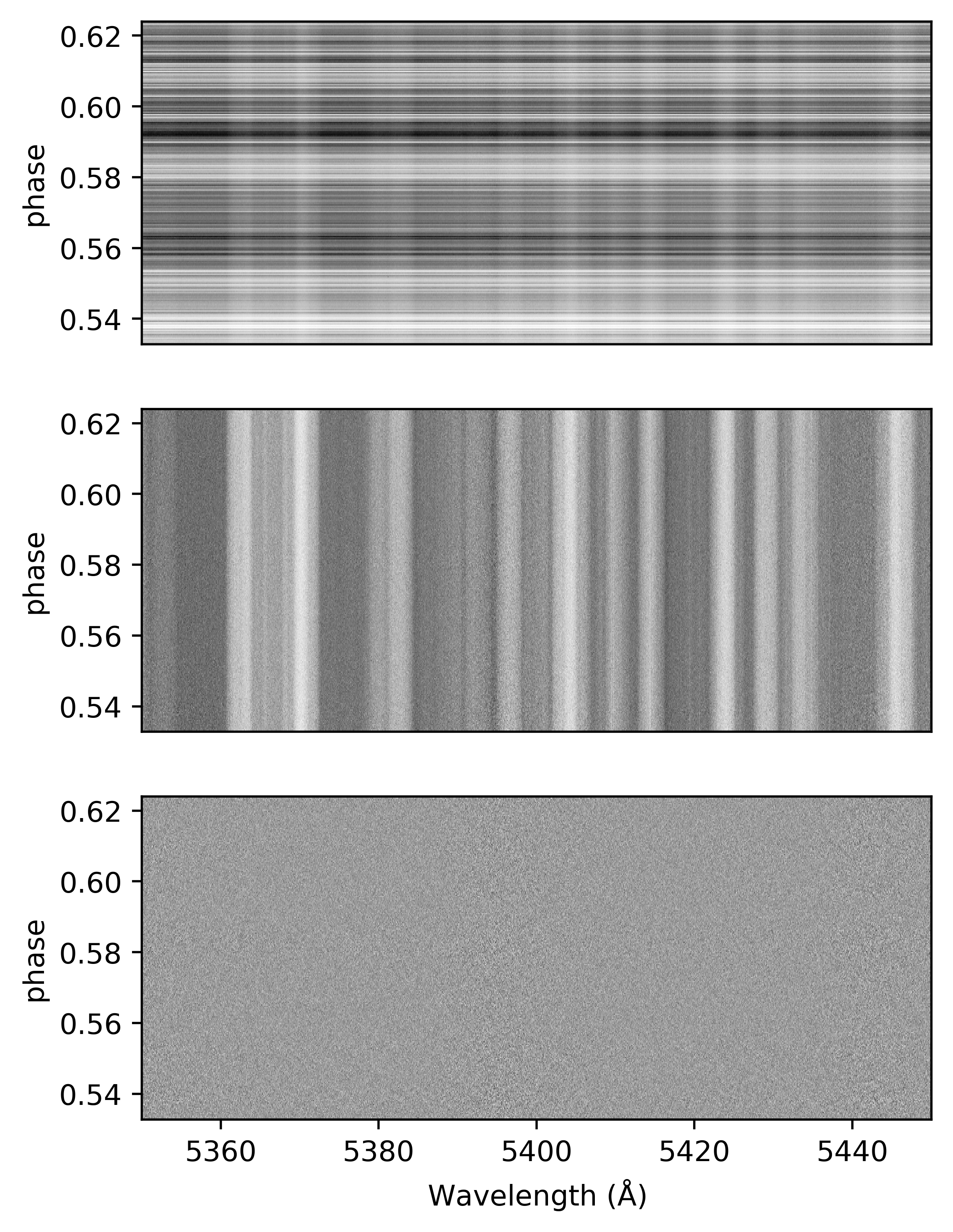}
      \caption{Example of the data-reduction procedures. These data are from Night-1 observation.
      \textit{(a)} Original spectra from the HARPS-N pipeline (Data Reduction Software). These are order-merged spectra. Here the spectra are shifted into the Earth's rest frame.
    \textit{(b)} Spectra after normalisation and correction of blaze variations.
     \textit{(c)} Residual spectra after the removal of stellar and telluric lines using \texttt{SYSREM}.
     }
         \label{fig-data-processing}
   \end{figure}

\section{Additional figures}
   \begin{figure}
   \centering
   \includegraphics[width=0.49\textwidth]{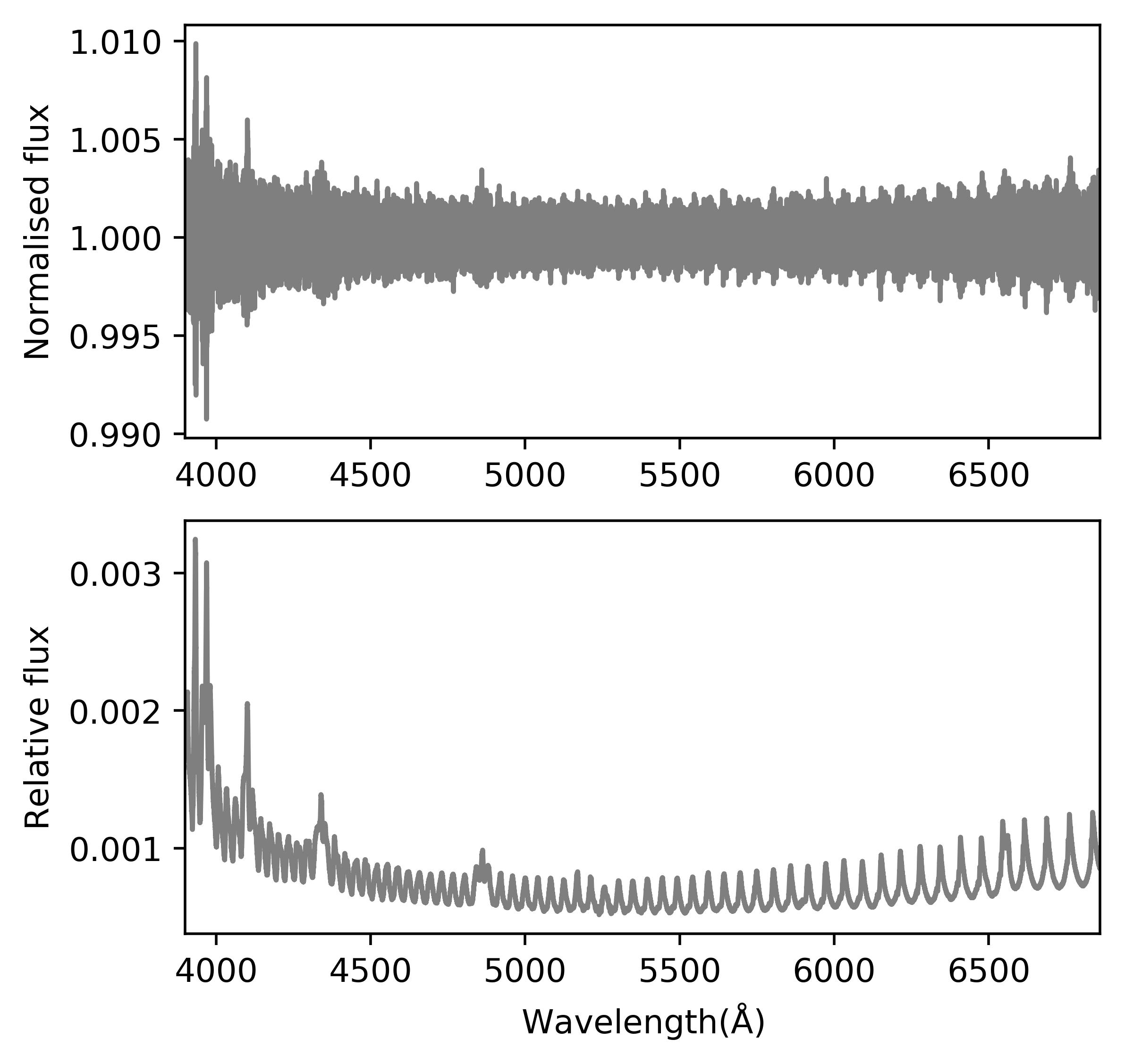}
      \caption{\textit{Upper panel:} Master residual spectrum. This is the combination of all the residual spectra of the two nights, excluding the spectra taken in eclipse.  \textit{Lower panel:} Uncertainty of the master residual spectrum.}
         \label{fig-master-residual}
   \end{figure}

   \begin{figure}
   \centering
   \includegraphics[width=0.49\textwidth]{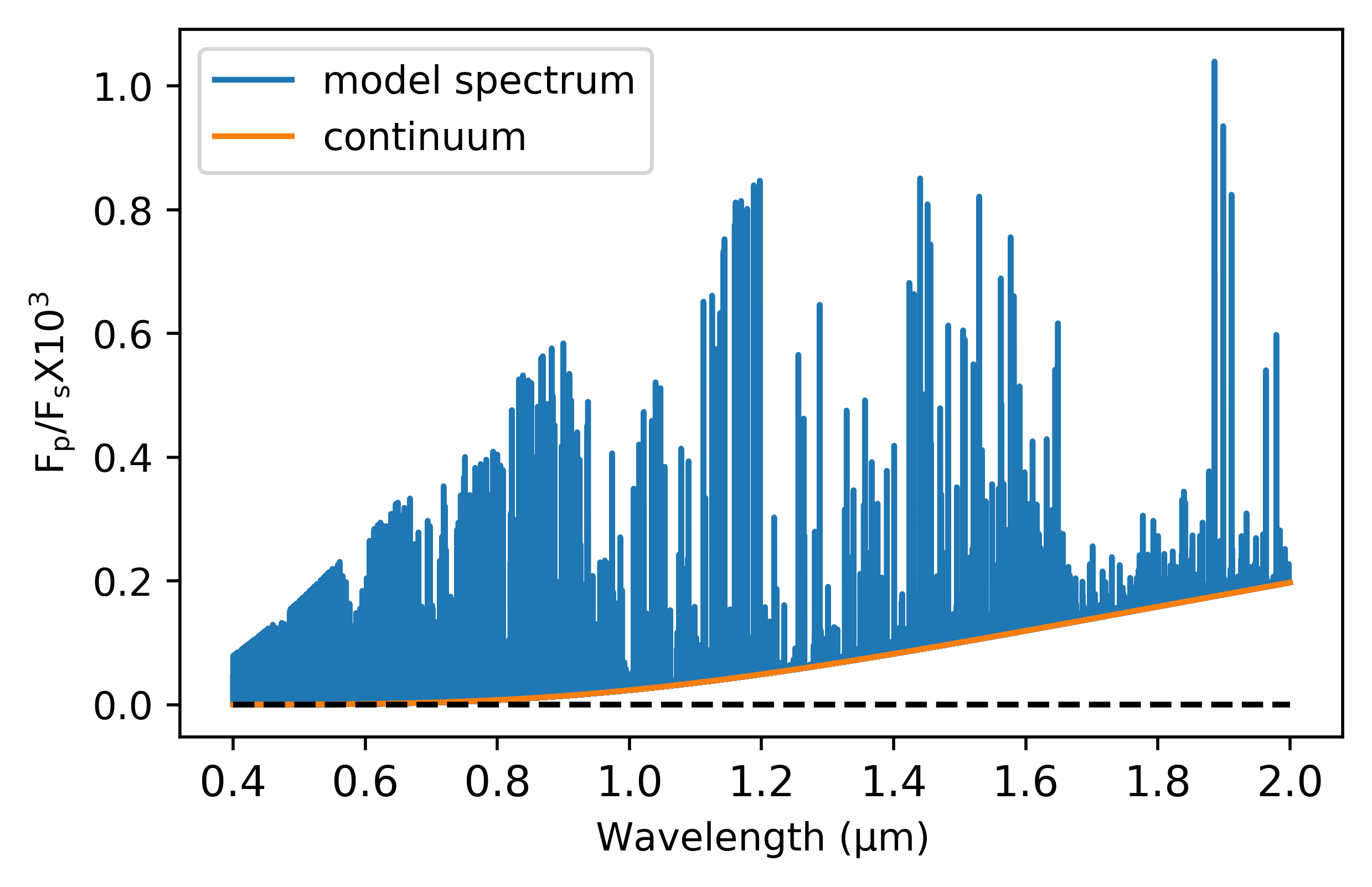}
      \caption{Example of the modelled spectrum (the blue line) and the corresponding continuum (the orange line). The dashed black line shows the zero flux level. The spectra were calculated using the best-fit $T$-$P$ profile in Fig.\ref{fig-best-TP}. The continuum corresponds to the blackbody spectrum at the temperature of $T_2$. In the HARPS-N wavelength range (i.e. below 0.69 $\mathrm{\mu m}$), the flux at the \ion{Fe}{i} line core is much larger than the continuum flux.}
         \label{fig-continuum}
   \end{figure}

   \begin{figure}
   \centering
   \includegraphics[width=0.49\textwidth, height=0.3\textwidth]{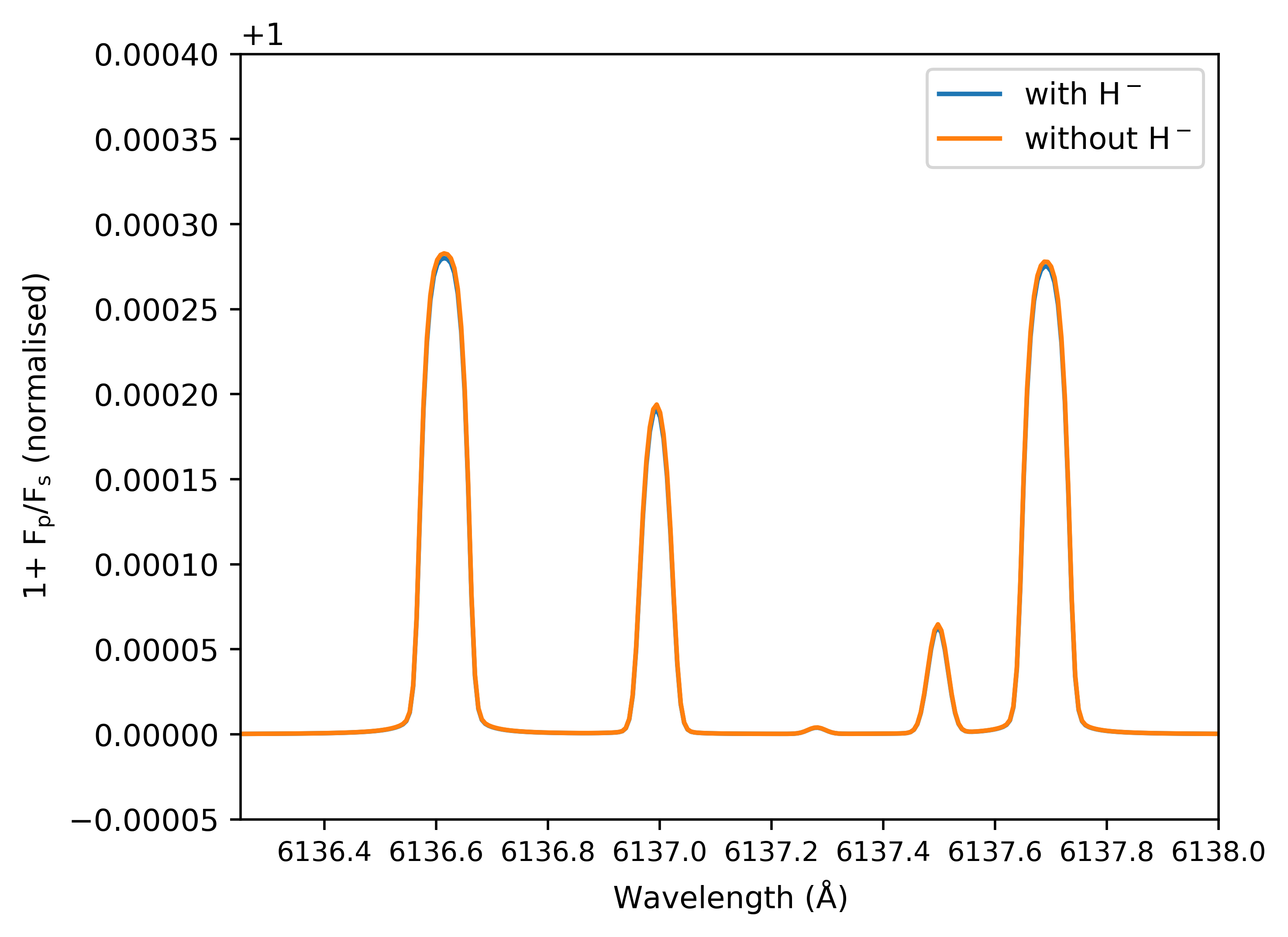}
      \caption{Comparison between the normalised model spectra with and without the $\mathrm{H^-}$ continuum opacity. The spectra were calculated using the same parameters as in Fig.\ref{fig-continuum}. The line depths between the two models are almost the same, indicating that the $\mathrm{H^-}$ contribution can be neglected in our analysis.}
         \label{fig-Hminus}
   \end{figure}

   \begin{figure}
   \centering
   \includegraphics[width=0.49\textwidth, height=0.3\textwidth]{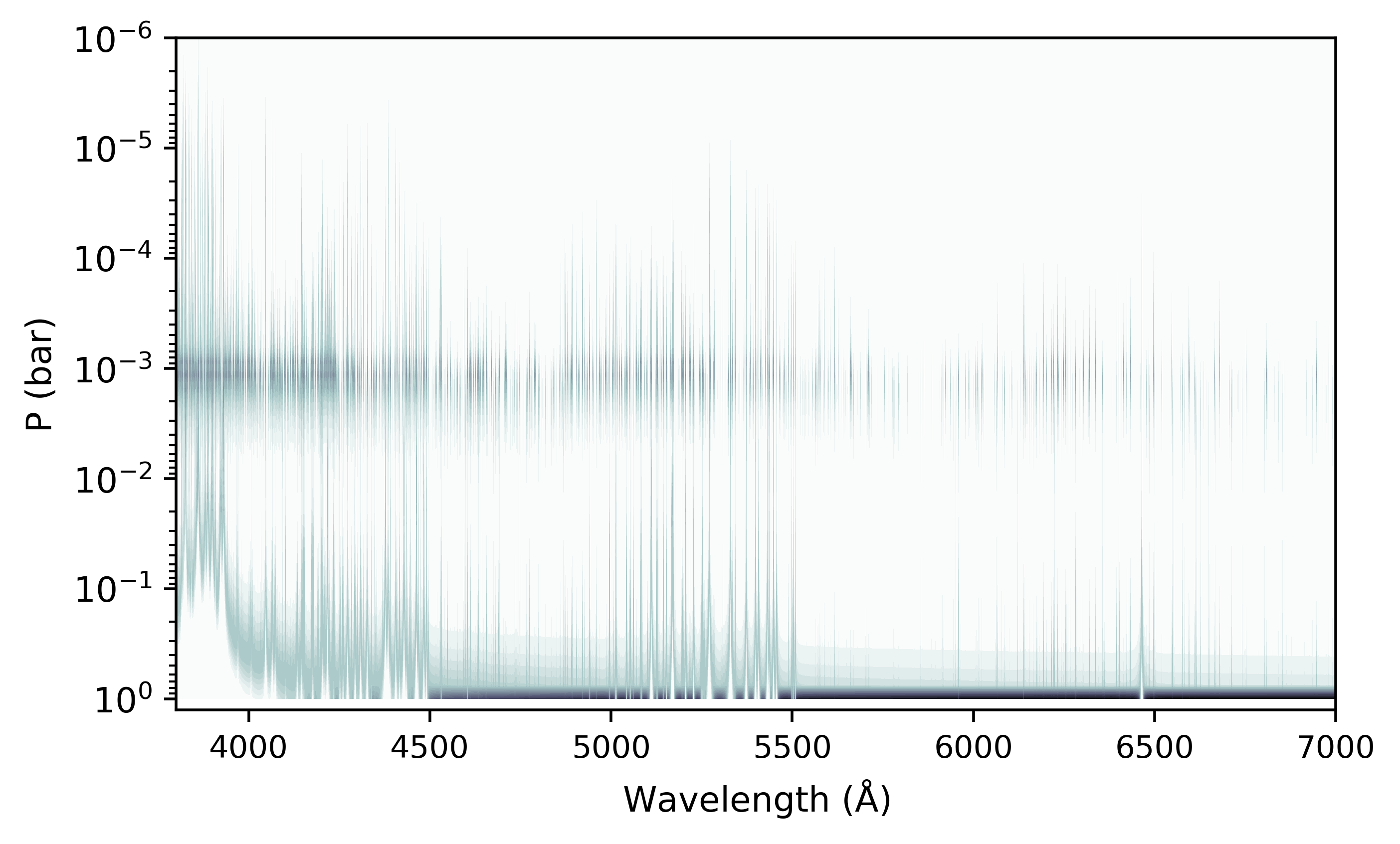}
   \includegraphics[width=0.49\textwidth, height=0.3\textwidth]{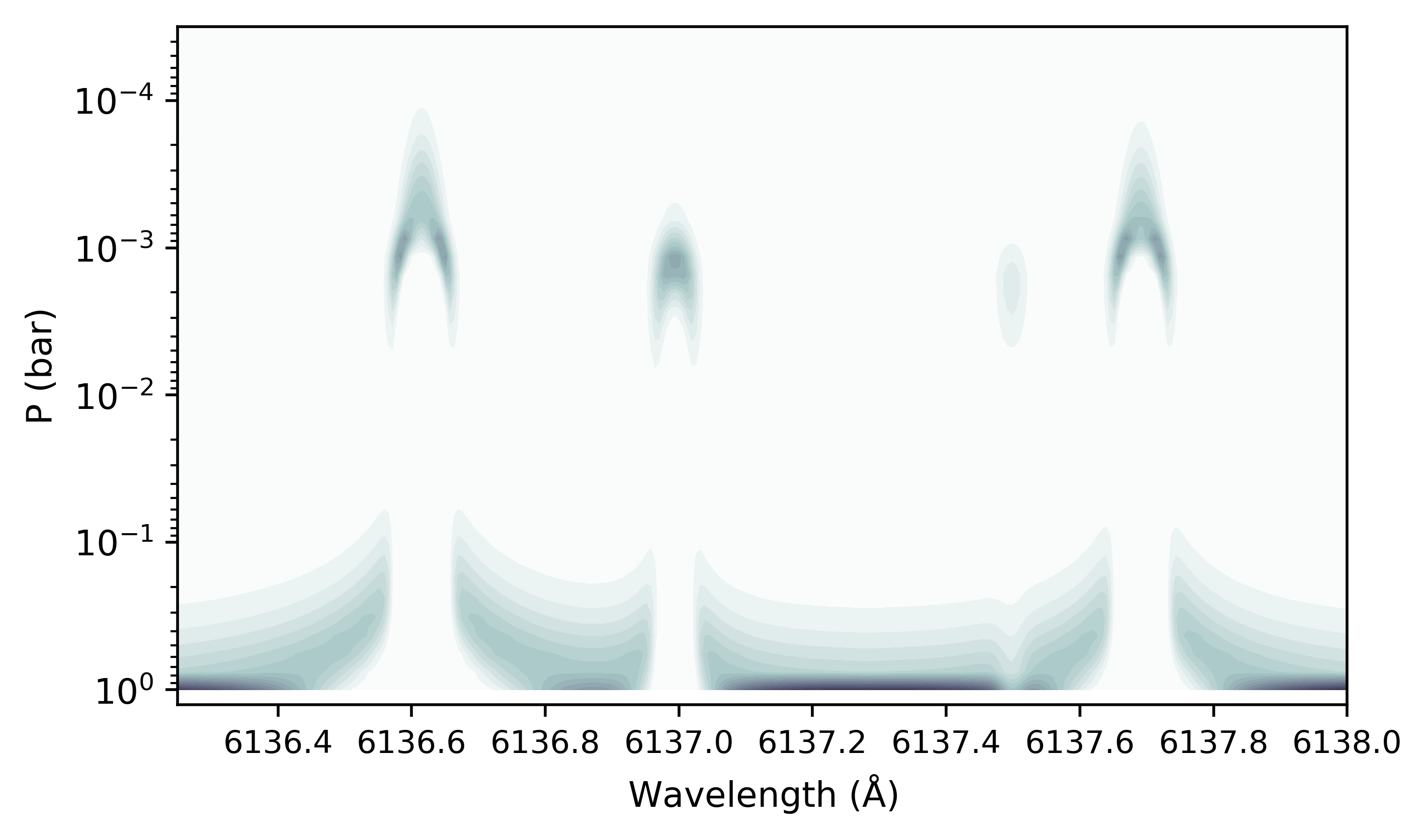}
      \caption{Contribution functions of the thermal emission spectrum calculated using the same parameters as in Fig.\ref{fig-continuum}. The upper panel presents the contribution functions over the entire HARPS-N wavelength range and the bottom panel presents a detailed view in a narrow wavelength range.   The contribution functions indicate that the fluxes of the Fe lines are from the upper layer of the temperature inversion, while the continuum flux is from the low altitudes.
      It appears that we are probing the continuum at the bottom of the atmosphere (which is set as 1 bar), but this result is due to the model setup (i.e. isothermal for pressures larger than $P_2$). The continuum spectrum is the same if we set the atmospheric lower boundary at $P_2$ instead of 1 bar. 
       }
         \label{fig-contribution}
   \end{figure}

   \begin{figure*}
   \centering
   \includegraphics[width=0.9\textwidth]{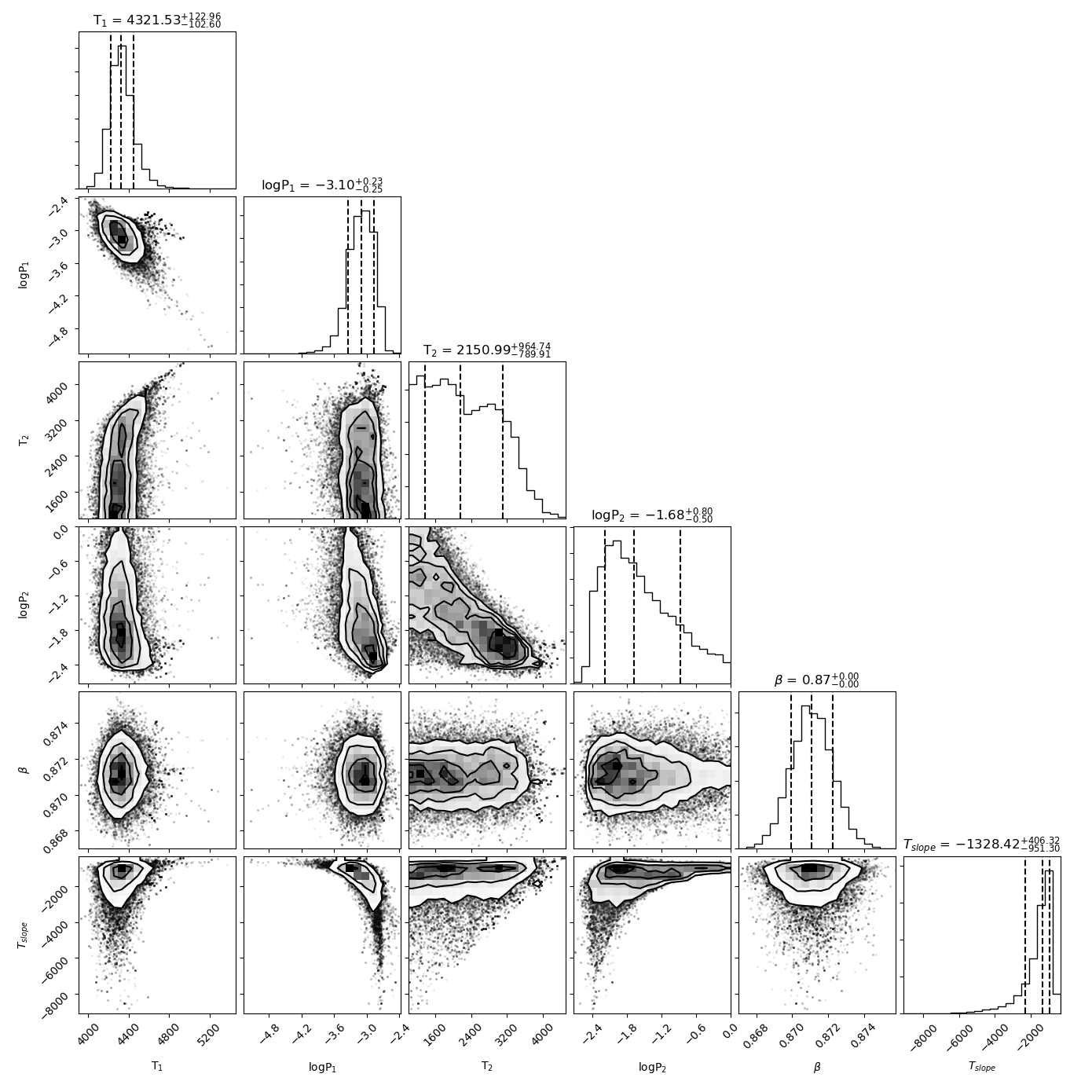}
      \caption{Posterior distribution of the parameters from the MCMC fit. $T_\mathrm{slope}$ is not a fitted parameter, but is calculated using Eq. (1).}
         \label{fig-MCMC-TP}
   \end{figure*}


\end{document}